
\nopagenumbers
\magnification=\magstep1
\parskip 0pt
\parindent 15pt
\baselineskip 16pt
\hsize 5.53 truein
\vsize 8.5 truein

\font\titolo = cmbx10 scaled \magstep2
\font\autori = cmsl10 scaled \magstep2
\font\abstract = cmr9

\hrule height 0pt
\vskip 0.5 truein
\rightline{POLFIS-TH.06/93}
\vskip 1.5 truecm
\centerline{\titolo MEAN FIELD RENORMALIZATION GROUP}
\vskip .05truein
\centerline{\titolo FOR THE BOUNDARY MAGNETIZATION}
\vskip .05truein
\centerline{\titolo OF STRIP CLUSTERS}
\vskip .15truein
\centerline{\autori Alessandro Pelizzola$^a$ and Attilio Stella$^b$}
\vskip .15truein
\centerline{\abstract $^a$Dipartimento di Fisica, Politecnico di Torino,
I-10129 Torino, Italy}
\centerline{\abstract $^b$Dipartimento di Fisica, Universit\'a di Bologna,
I-40126 Bologna, Italy}
\vskip .35 truein
\centerline{\bf ABSTRACT}
\vskip .05truein {\abstract We analyze in some detail a recently proposed
transfer matrix mean field approximation which yields the exact critical
point for several two dimensional nearest neighbor Ising models. For the
square lattice model we show explicitly that this approximation yields
not only the exact critical point, but also the exact boundary
magnetization of a semi--infinite Ising model, independent of the size of
the strips used. Then we develop a new mean field renormalization group
strategy based on this approximation and make connections with finite size
scaling. Applying our strategy to the quadratic Ising and
three--state Potts models we obtain results for the critical exponents
which are in excellent agreement with the exact ones. In this way we also
clarify some advantages and limitations of the mean field renormalization
group approach.}
\par
\vskip .35 truein
\vfill
\eject

\pageno=1
\footline{\hss\tenrm\folio$\,$\hss}
\def\makefootline{\baselineskip=50pt\line{\the\footline}}
\parindent 15pt

{\bf 1. Introduction} \par \medskip
A recently proposed[1] transfer matrix version of a mean field approximation
(which in the following will be denoted by LS) applied to several nearest
neighbor Ising models in two dimensions, gave surprisingly exact results
for the critical points, even without extrapolation, and very good results,
under extrapolation, for more complicated models. \par
A first issue to address in connection with the LS approximation is why the
results are exact in the n.n. Ising case and extrapolate accurately in the
others, and whether extra exact results can be obtained by this scheme. \par
We want also to clarify what is the connection of this method with other
techniques of more common use in two dimensional statistical mechanics. In
particular:
\item{ a) } since the method involves consideration of strips similar to
those used, e.g., in finite size scaling (FSS) and phenomenological
renormalization approaches[2], it is legitimate to ask up to which extent the
LS approximation is connected to these approaches and possibly fits within
them;
\item{ b) } since the method uses as a basic ingredient effective
fields on the boundary of the strips, it is also rather natural to look for
connections with the so--called mean field renormalization group[3] (MFRG)
approach. This will go together with showing how critical exponents can be
obtained in this context. \par
The plan of the paper is as follows: in Sec. 2 we give a brief review of
the LS approximation and show, in the square lattice Ising case, that it
gives not only the exact critical point, but also the exact boundary
magnetization of a semi--infinite Ising model, independent of the size of
the strips used; in Sec. 3 we show how the LS approximation fits into a
MFRG structure, develop a procedure for calculating the critical exponents,
and compare the method with the FSS approach; in Sec. 4 we give two test
applications of the procedure above, on the two dimensional Ising and
three--state Potts models. Finally, in Sec. 5, we draw some conclusions. \par
\vfill\eject
{\bf 2. The LS Approximation} \par
\medskip
The LS approximation scheme[1]
makes use of two infinite strips ${\cal S}_n$ and ${\cal
S}_{n^\prime}$ of widths $n$ and $n^\prime$ respectively, with
periodic boundary conditions along the infinite direction.
The approximation is obtained by applying an effective field $h_{eff}$
at {\it one side} of the strips and by imposing the consistency relation
$$m_{1 n}(K,h_{eff}) = m_{1 n^\prime}(K,h_{eff}),\eqno(1)$$
where $m_{1 n}$ and $m_{1 n^\prime}$ are the values of the order parameter at
{\it the opposite side} of the strips and $K = \beta J > 0$ is the exchange
interaction strength. Eq. (1) has to be solved for $h_{eff}$ at fixed $K$
and the critical temperature is the one for which the paramagnetic
solution $h_{eff} = 0$ bifurcates into nonzero solutions, leading to
spontaneous magnetization. \par
This method, as pointed out by Lipowski and Suzuki[1], yields the exact
critical temperature of the Ising model with nearest--neighbor interaction
on many two dimensional lattices (square, triangular, honeycomb and
centered square), and very accurate estimates of the critical temperature
of more complicated models (Ising model with alternating strength of
interaction, with next--nearest--neighbor interaction, $S \ge 1$ models). \par
In the present section we will show, resorting to a result by Au--Yang and
Fisher[4], that at least in the simplest case of the nearest--neighbor
Ising model on the square lattice the LS approximation yields not only the
exact
critical temperature, but also the {\it exact boundary magnetization} of
the semi--infinite model. \par
Let us consider an Ising model on a strip ${\cal S}_n$ of a square lattice,
number from $1$
to $n$ the chains which form the strip and apply a magnetic field $h_n$ on
the $n$th chain. The corresponding hamiltonian will be
$$- \beta H = K \sum_{i = - \infty}^{+ \infty} \sum_{j = 1}^{n - 1}
(s_{i,j} s_{i+1,j} + s_{i,j} s_{i,j+1}) +
K \sum_{i = - \infty}^{+ \infty} s_{i,n} s_{i+1,n} +
h_n \sum_{i = - \infty}^{+ \infty} s_{i,n}, \eqno(2)$$
where $s_{i,j} = \pm 1$
is an Ising spin located at the site with coordinates $i$ and $j$ in the
$x$ and $y$ direction respectively. The magnetization $m_{1 n} \equiv
m_{1 n}(K,h_n) = \langle s_{i,1} \rangle$ has been calculated for $n \ge 2$
by Fisher and
Au--Yang in ref. [4], and is given by
$$m_{1 n}(K,h_n) = z^\prime \left( {\tilde c_+ \over \tilde c_-} \right)^{1/2}
\left[ {\vert \tilde t \vert - \tilde t \ {\rm tanh}(2 n \ {\rm sinh}^{-1}
\vert t^\prime \vert ) \over
\vert \tilde t \vert + (\tilde t + \tilde c_+ {z^{\prime}}^2 )
{\rm tanh}(2 n \ {\rm sinh}^{-1} \vert t^\prime \vert ) } \right]^{1/2}.
\eqno(3)$$
In eq. (3) we have adopted the same notation of ref. [4], i.e.
$$z^\prime = {\rm tanh} h_n,$$
$$t^\prime = {1 - {\rm sinh} 2K \over (2 {\rm sinh} 2K)^{1/2}}, \qquad
\tilde t = t^\prime(1 + {t^\prime}^2)^{1/2} = {1 \over 2}({\rm coth} 2K -
{\rm cosh} 2K),$$
$$\tilde c_\pm = (2 + {t^\prime}^2)^{1/2} (1 + {t^\prime}^2)^{1/2} - \tilde
t \pm 1 = {\rm cosh} 2K \pm 1.$$
For $K > K_c = {1 \over 2}\ln(1 + \sqrt{2})$
it is easily realised that if $h_n$ is chosen in such a way that
$\tilde t + \tilde c_+ {z^{\prime}}^2 = - \tilde t$, i.e.
$$z^\prime = {\rm tanh} h_n = \left( {{\rm cosh} 2K - {\rm coth} 2K \over
{\rm cosh} 2K + 1} \right)^{1/2} \eqno(4)$$
then the quantity in
square brackets in eq. (3) equals 1, independent of $n$, and one has
$$m_{1 n} (K, h_n) \equiv m_1(K) = \left( {{\rm cosh} 2K - {\rm coth} 2K \over
{\rm cosh} 2K - 1} \right)^{1/2} \eqno(5)$$
where $m_1(K)$ is the {\it exact boundary magnetization} of the two
dimensional semi--infinite Ising model[5]. Furthermore, for $K < K_c$,
choosing $h_n = 0$ yields $m_{1}(K) = 0$, again independent on $n$. \par
The case $n = 1$, for which eq. (3) is no more valid, can be easily solved by
the transfer matrix method, obtaining
$$m_{1 1}(K,h_1) = {{\rm sinh} h_1 \over \left( {\rm sinh}^2 h_1 + e^{-4K}
\right)^{1/2}},\eqno(6)$$
which again equals $m_1(K)$ if $h_1$ is chosen according to eq. (4). \par
These results imply that, no matter which $n,n^\prime$ we choose, the
bifurcation at $h_{eff} = 0$ will always occur at the exact critical
point, $K = K_c$. This mechanism is at the basis of the results of ref. [1]
for the square lattice case, and we believe that the same should work for
the other two dimensional lattices. \par
Finally, if, fully in the spirit of a classical approach, we consider the
nonzero magnetization solution in eq. (1) for $K > K_c$, by putting
$h_{eff} = h_n$ we obtain the exact spontaneous boundary magnetization $m_1$,
independent of $n,n^\prime$. This was not noticed in ref. [1]. \par
The following remarks are in order for an explanation of the above results.
First of all, the boundary magnetization is known to behave as $m_1(K)
\approx (K - K_c)^{\beta_1}$ for $K \to K_c^+$ with $\beta_1 = 1/2$, in the
$2d$ Ising model. The exponent $\beta_1 = 1/2$ [5] is such that it can be
reproduced exactly by a mean field like approximation, making use of an
effective field. When considering models with $\beta_1$ values incompatible
with a classical scheme, one has to consider the LS approach and its
possible extensions and approximations, as we will discuss in the next
sections. \par
As shown in ref. [6], in the context of a generalized cluster variation
approach to two dimensional lattice models, the double strip ${\cal S}_2$
is able to contain the whole information needed to solve exactly the two
dimensional n.n. Ising model. The problem then reduces to how such
information can be extracted. Clearly what we presented in this section
amounts to a relatively simple way of obtaining part of this information. \par
\bigskip
{\bf 3. Mean field renormalization group and finite size scaling} \par
\medskip
Let us now see how the LS approximation can be used to develop a new MFRG
strategy, where the boundary magnetization is used together with the bulk
one as an effective scaling operator.
This will also be useful in understanding the relations of the LS
method with FSS. \par
The notation applies to an Ising model for convenience, but the strategy is
not limited to this case, as will be shown in the next section where it
will be used to investigate the three--state Potts model. \par
For infinite Ising strips of widths $n$ and $n^\prime$, FSS implies the
following scaling law
for the singular part of the bulk free energy density, $f^{(b)}$,
$$f_{n^\prime}^{(b)}(\ell^{y_T} \epsilon, \ell^{y_H} h) =
\ell^d f_n^{(b)}(\epsilon,h), \eqno(7)$$
where $\epsilon = (T_c - T)/T_c$, $\ell = n/n^\prime$ is the rescaling factor,
and $d$ is the bulk dimension. If the boundary conditions are open, for the
singular part of the surface free energy density, $f^{(s)}$, the relation
$$f_{n^\prime}^{(s)}(\ell^{y_T} \epsilon, \ell^{y_H} h, \ell^{y_{HS}} b) =
\ell^{d - 1} f_n^{(s)}(\epsilon,h,b), \eqno(8)$$
holds, where $b$ indicates a surface field. \par
The MFRG basic idea[3] is to derive from eq. (7) the scaling relation for the
bulk
magnetization
$$m_{n^\prime}(K^\prime,h^\prime) = \ell^{d - y_H}m_n(K,h), \eqno(9)$$
where $h^\prime = \ell^{y_H}h$ and $K^\prime \equiv K^\prime(K)$ is a
mapping in the Wilson--Kadanoff sense, determined implicitly, in the limit
of $h$ going to zero, on the basis of eq. (9).
{}From this mapping the critical point $K_c$ and the thermal exponent $y_T$ can
be
obtained by means of the relations $K_c = K^\prime(K_c)$ and $\ell^{y_T} =
\displaystyle{\partial K^\prime \over \partial K}(K = K_c)$. \par
Applying this idea to the surface magnetization yields
$$m_{1 n^\prime}(K^\prime, h^\prime, b^\prime) =
\ell^{d - 1 - y_{HS}}m_{1 n}(K,h,b),\eqno(10)$$
where $b^\prime = \ell^{y_{HS}}b$ if we want $b$ to scale as a surface
field. On the other hand, the equation for the critical point which is
characteristic of the LS approximation would be recovered if $b$ scaled as
a magnetization, i.e. with an exponent $d - 1 - y_{HS}$. In fact, with this
assumption, setting $h = 0$ and linearizing in $b$ yields
$${\partial m_{1 n^\prime} \over \partial b^\prime} (K^\prime,0,0) =
{\partial m_{1 n} \over \partial b} (K,0,0), \eqno(11)$$
which implicitly defines a mapping $K^\prime(K)$. The equation
$$K_c = K^\prime(K_c)\eqno(12)$$
with $K^\prime(K)$ given by eq. (11) is equivalent to the equation for the
critical point in the LS approximation.
So this approximation can also be seen as a realization of a MFRG strategy
as far as determination of $K_c$  is concerned. \par
In a MFRG spirit one can also determine the critical exponents, since $y_T$ is
obtained by the relation
$$\ell^{y_T} = \left. {\partial K^\prime \over \partial K}
\right\vert_{K=K_c}, \eqno(13)$$
linearizing eq. (9) with respect to $h$, with $K = K_c$ yields
$${\partial m_{n^\prime} \over \partial h^\prime} (K_c,0) = \ell^{d - 2 y_H}
{\partial m_n \over \partial h} (K_c,0), \eqno(14)$$
from which $y_H$ can be obtained and finally, linearizing eq. (10) in the same
way, with $b = 0$, yields
$${\partial m_{1 n^\prime} \over \partial h^\prime} (K_c,0,0) = \ell^{d - 1 -
y_{HS} - y_H} {\partial m_{1 n} \over \partial h} (K_c,0,0), \eqno(15)$$
from which $y_{HS}$ is obtained. \par
The set of equations (11)--(15) is a MFRG procedure to determine critical point
and critical exponents. \par
Nevertheless, the procedure above (to be denoted by M, for
''magnetization'', in the following) is
not a rigorous application of FSS. In such an application (11) should be
replaced by
$${\partial m_{1 n^\prime} \over \partial b^\prime} (K^\prime,0,0) =
\ell^{d - 1 - 2 y_{HS}}{\partial m_{1 n} \over \partial b} (K,0,0), \eqno(16)$$
since $b$ should scale as a field, with exponent $y_{HS}$, and should be
solved in conjunction with (14)--(15). This alternative and more rigorous
procedure will be denoted by F, for ''field'', in the following. \par
F is in fact the procedure of MFRG proposed in ref. [3] to yield
simultaneously bulk and surface exponents. It is interesting to
investigate how M, proposed here, being more consistent with the effective
field idea, compares with F. \par
Two comments are in order:
\item{ i) } the two procedures should give the same value of $K_c$ (but not
of the exponents) in the limit $n,n^\prime \to \infty$, $\ell \to 1$, since
the two derivatives in eq. (11) are analytic functions; this should justify
the LS approximation in a FSS context;
\item{ ii) } in the two dimensional n.n. Ising case, $y_{HS} = 1/2$ exactly
and then, in the limit above, also the critical exponents should be the
same for both procedures M and F. \par
In the following section we will check these ideas on the two dimensional
n.n. Ising and three--state Potts cases. \par
\bigskip
{\bf 4. Results and discussion} \par
\medskip
In the present section we give two test applications of our new MFRG
strategy (M),
to the Ising and three--state Potts models on square lattices. We also
compare our results with those obtained treating $b$ as a surface field,
i.e. letting it scale with exponent $y_{HS}$.
We start by applying procedure M to the Ising model. In the Ising
case, we have already shown that the method gives the exact critical point
for any $n$, $n^\prime$. Furthermore, resorting to eq. (3), the mapping
$K^\prime = K^\prime(K)$ can be determined analytically in an implicit
form. As a result one gets
$$f_{n^\prime}(K^\prime) = f_n(K), \eqno(17)$$
where
$$f_n(K) = K \ {\rm coth}K \left[ {1 + {\rm tanh}(2n \ {\rm sinh}^{-1}
\vert t^\prime \vert) \over 1 - {\rm tanh}(2n \ {\rm sinh}^{-1}
\vert t^\prime \vert)} \right]^{1/2} \eqno(18)$$
and with $t^\prime$ as above. It can be checked that the fixed
point of (17), obtained by setting $t^\prime = 0$, is $K^* = {1 \over 2}
\ln(1 + \sqrt{2})$, while for the thermal exponent one has
$$\ell^{y_T} = { 1 + 2(2n - 1)K^* \over 1 + 2(2n^\prime - 1)K^* },\eqno(19)$$
which, in the limit $n, n^\prime \to \infty$, yields $y_T = 1$, which is
again an exact result. \par
The calculation of the magnetic exponents cannot be carried out
analytically since no solution is available for the bulk magnetization of a
strip in the presence of a bulk magnetic field, and we have to proceed
numerically, as follows. Given the strip ${\cal S}_n$ with a bulk magnetic
field $h$ and an auxiliary magnetic field $h_1$ acting on the first chain,
we determine its partition function $Z_n(K,h,h_1)$ as the largest
eigenvalue of the $2^n \times 2^n$ transfer matrix with elements
$$\eqalign{
T_n(\{s_{j}\},\{s_{j}^\prime\}) = \exp &
\left[ {K \over 2} \sum_{j=1}^{n-1} (s_j s_{j+1} + s_j^\prime s_{j+1}^\prime)
+ K \sum_{j=1}^n s_j s_j^\prime \right. \cr
& \left. + {h \over 2} \sum_{j=1}^n (s_j + s_j^\prime)
+ {h_1 \over 2} (s_1 + s_1^\prime) \right]. \cr} \eqno(20)$$
The bulk and boundary magnetizations will then be given by
$$m_n \equiv m_n(K,h) = \left. {1 \over Z_n} {\partial Z_n \over \partial
h} \right\vert_{h_1 = 0} \eqno(21)$$
and
$$m_{1 n} \equiv m_{1 n}(K,h) = \left. {1 \over Z_n}
{\partial Z_n \over \partial h_1} \right\vert_{h_1 = 0} \eqno(22)$$
respectively. Finally $y_H$ and $y_{HS}$ are determined according
to (14)--(15). \par
In Tab. 1 we report the results for the critical exponents for strip widths
$2 \le n = n^\prime + 1 \le 8$ ($n = n^\prime + 1$ is always the most
convenient choice). The extrapolations are based on least squares fits with
fourth order polynomials in $1/n$ and are certainly justified since the
critical exponents are nearly linear functions of $1/n$.
The agreement of the extrapolated data with the exact results is
very good, and the errors are within $0.2\%$. \par
The Ising test has yielded very promising results, but does not shed much
light on the physical meaning of the effective parameter $b$, since for the
two dimensional Ising model surface magnetization and surface field scale
with the same exponent $d - 1 - y_{HS} = y_{HS} = 1/2$. In fact, procedure
F, in which $b$ scales
as a surface field yields as well very good results (apart from having no
solution when $n = 2$), as shown in Tab. 2.
Data from the two procedures are plotted together in Figs. 1--4.
As expected, all the results seem to be equivalent in
the limit $n \to \infty$. \par
In view of the above considerations, we believe that a
more conclusive test is in order, and a suitable model should be the
three--state Potts model. Indeed, in two dimensions, this model
is known to undergo a second order phase transition, whose critical point
and critical exponents are known exactly [7,8], even in the absence of a full
solution. In particular it has $y_{HS}\ne 1/2$. \par
The hamiltonian of the $q$--state Potts model[7] is
$$- \beta H = {K \over q - 1} \sum_{\langle i j \rangle}
(q \delta_{s_i,s_j} - 1) + {h \over q - 1} \sum_i
(q \delta_{s_i,0} - 1), \eqno(23)$$
where the variables $s_i$ take on values $0,1, \ldots q - 1$, $K > 0$ is
the interaction strength and $h$ is a magnetic field. The order
parameter of the model, corresponding to the Ising magnetization, is
$$m = {q \langle \delta_{s_i,0} \rangle - 1 \over q - 1}. \eqno(24)$$
In the case $q = 2$ one recovers the Ising model. \par
The MFRG scheme developed above can be carried over to the $q$--state Potts
model without any substantial modification, and we will apply it to the
case $q = 3$. The main new fact is that no analytical results like eq. (3) are
available for the three--state Potts model. So all calculations must be
performed numerically with the transfer matrix method. The order of the
transfer matrix is now $3^n$ and increases more rapidly than in the Ising
case. However, the transfer matrix is invariant with respect to the
transformation which interchanges the states $s_i = 1$ and $s_i = 2$ (all
other symmetries are lost as soon as one introduces the surface fields),
and the eigenvector corresponding to its largest eigenvalue belongs to the
symmetric subspace of this transformation. Thus we can limit ourselves to
matrices acting in this subspace, which are of order $(3^n + 1)/2$. In this
way we have been able to deal with strips up to $n = 7$. \par
The numerical results for procedure M
are reported in Tab. 3. The extrapolation is obtained by
fitting data in a least square sense with a second order polynomial in $1/n$.
Even if now the critical point is not given exactly by the LS
approximation, we
obtain excellent agreement with the exact results (error within
2.3\%) for the bulk critical point and exponents, while, in comparison, the
results for the
surface exponent $y_{HS}$ are rather poor. \par
In Tab. 4 we report the results obtained from procedure F.
The situation is very similar to the previous one: the errors of the
extrapolated critical point and bulk exponents are within 4.7\% and, in
comparison, the estimate of the surface exponent is again rather poor.
Results from the two procedures are plotted together in Figs. 5--8. \par
There is some evidence that F is the correct procedure when $n$ is
large, but for small strips M, although not rigorous, seems to
work very well: indeed if $y_H$ had been extrapolated on the basis of the
results for $2 \le n \le 5$ one would have obtained $1.868$, which is two
orders of magnitude more accurate than the extrapolation on the whole set
of data. \par \bigskip
{\bf 5. Conclusions} \par
\medskip
We have analyzed in some detail the LS approximation, showing that in the
two dimensional n.n. Ising case, it yields not only the exact critical
point, but also the exact boundary magnetization of the semi--infinite
model, independent of the size of the strips used. We have also proposed an
explanation of these surprising results. \par
The LS approximation has been used to develop a new MFRG strategy
(procedure M) which yielded very accurate results for the critical
exponents of the Ising and three--state Potts models in two dimensions.
When compared with rigorous FSS (procedure F) our new strategy has proven
to be particularly suitable for applications where relatively small strips
are used. \par
\vfill\eject
{\bf References} \par \medskip
\item{[1]} A. Lipowski and M. Suzuki, J. Phys. Soc. Jpn. {\bf 61} (1992) 4356.
\item{[2]} M.N. Barber, in {\it Phase Transitions and Critical Phenomena},
edited by C. Domb and  J.L. Lebowitz (Academic, Lonodon, 1983), vol. 8.
\item{[3]} J.O. Indekeu, A. Maritan and A.L. Stella, Phys. Rev. B {\bf 35}
(1987) 305; for a review see K. Croes and J.O. Indekeu, preprint Katholieke
Universiteit Leuven (1993).
\item{[4]} H. Au--Yang and M.E. Fisher, Phys. Rev. B {\bf 21} (1980) 3956
(here our eq. (3) is reported with a misprint).
\item{[5]} B.M. McCoy and T.T. Wu, {\it The Two--Dimensional Ising Model}
(Harvard University Press, Cambridge, Mass., 1973), Chap. 6.
\item{[6]} A.G. Schlijper, J. Stat. Phys. {\bf 35} (1984) 285.
\item{[7]} F.Y. Wu, Rev. Mod. Phys. {\bf 54} (1982) 235.
\item{[8]} J.L. Cardy, Nucl. Phys. B {\bf 240} (1984) 514.
\vfill\eject
{\bf Figure Captions}\par \medskip
\parindent .4 truein
\item{}
\itemitem{\hbox to .67 truein{Fig. 1 : \hfill}} The Ising critical point
$K_c$ vs. $1/n$ as given by the M (solid line) and F (dashed line) procedures.
\itemitem{\hbox to .67 truein{Fig. 2 : \hfill}} The Ising thermal exponent
$y_T$ vs. $1/n$ as given by the M (solid line) and F (dashed line) procedures.
\itemitem{\hbox to .67 truein{Fig. 3 : \hfill}} The Ising magnetic exponent
$y_H$ vs. $1/n$ as given by the M (solid line) and F (dashed line) procedures.
\itemitem{\hbox to .67 truein{Fig. 4 : \hfill}} The Ising surface magnetic
exponent
$y_{HS}$ vs. $1/n$ as given by the M (solid line) and F (dashed line)
procedures.
\itemitem{\hbox to .67 truein{Fig. 5 : \hfill}} The three--state Potts critical
point
$K_c$ vs. $1/n$ as given by the M (solid line) and F (dashed line) procedures.
\itemitem{\hbox to .67 truein{Fig. 6 : \hfill}} The three--state Potts thermal
exponent
$y_T$ vs. $1/n$ as given by the M (solid line) and F (dashed line) procedures.
\itemitem{\hbox to .67 truein{Fig. 7 : \hfill}} The three--state Potts magnetic
exponent
$y_H$ vs. $1/n$ as given by the M (solid line) and F (dashed line) procedures.
\itemitem{\hbox to .67 truein{Fig. 8 : \hfill}} The three--state Potts surface
magnetic exponent
$y_{HS}$ vs. $1/n$ as given by the M (solid line) and F (dashed line)
procedures.
\vfill \eject
\nopagenumbers
\obeyspaces
\parindent 0pt
\tt
Tab. 1: results for the Ising model ($b = $ magnetization) \par
\medskip

$n$          $y_T$       $y_H$       $y_{HS}$    \par
------------------------------------ \par
2        0.95738  1.60287  0.60287  \par
3        0.97310  1.68092  0.56808  \par
4        0.98089  1.72197  0.55099  \par
5        0.98514  1.74793  0.54067  \par
6        0.98785  1.76600  0.53374  \par
7        0.98969  1.77937  0.52873  \par
8        0.99106  1.78970  0.52496  \par
------------------------------------ \par
Extrap   0.99977  1.87201  0.49764  \par
------------------------------------ \par
Exact    1        1.875    0.5      \par
\vskip 2 truecm

Tab. 2: results for the Ising model ($b = $ field) \par
\medskip

$n$          $K_c$       $y_T$       $y_H$       $y_{HS}$    \par
--------------------------------------------- \par
3        0.52903  0.96548  1.99012  0.91150  \par
4        0.48065  0.97840  1.91195  0.77040  \par
5        0.46366  0.98404  1.88570  0.70267  \par
6        0.45560  0.98726  1.87397  0.66195  \par
7        0.45113  0.98938  1.86794  0.63455  \par
8        0.44839  0.99086  1.86461  0.61482  \par
--------------------------------------------- \par
Extrap   0.44512  0.99858  1.87440  0.49865  \par
--------------------------------------------- \par
Exact    0.44069  1        1.875    0.5      \par
\vfill \eject

Tab. 3: results for the 3--state Potts model ($b = $ \par magne\-tization) \par
\medskip

$n$          $K_c$       $y_T$       $y_H$       $y_{HS}$ \par
--------------------------------------------- \par
2        0.71318  1.10152  1.60916  0.60916  \par
3        0.69762  1.12118  1.69079  0.56966  \par
4        0.69013  1.13209  1.73652  0.54785  \par
5        0.68574  1.13978  1.76704  0.53350  \par
6        0.68287  1.14496  1.78923  0.52317  \par
7        0.68084  1.14993  1.80629  0.51526  \par
--------------------------------------------- \par
Extrap   0.66896  1.17561  1.90880  0.47250  \par
---------------------------------------------- \par
Exact    0.67004  1.2      1.86667  0.33333  \par
\vskip 2 truecm

Tab. 4: results for the 3--state Potts model ($b = $ field) \par
\medskip

$n$          $K_c$       $y_T$       $y_H$       $y_{HS}$    \par
--------------------------------------------- \par
3        0.78999  1.12871  1.96588  0.87340  \par
4        0.72612  1.14060  1.88767  0.72035  \par
5        0.70302  1.14627  1.86166  0.64290  \par
6        0.69184  1.15025  1.85030  0.59417  \par
7        0.68555  1.15541  1.84477  0.56018  \par
--------------------------------------------- \par
Extrap   0.70167  1.17551  1.88526  0.40703  \par
--------------------------------------------- \par
Exact    0.67004  1.2      1.86667  0.33333  \par

\vfill\eject\end